\documentclass[usenatbib,referee]{biom}
\usepackage{color}
\usepackage{amssymb}
\usepackage{amsmath}
\usepackage{amsfonts}
\usepackage{graphicx}
\usepackage{caption}
\usepackage{float}
\usepackage{booktabs}
\usepackage{multirow}
\usepackage{subfigure}
\usepackage{setspace}
\usepackage{endfloat}

\newtheorem{condition}{Condition}

\newcommand{\tl}[1]{\multicolumn{1}{l}{#1}} 
\newcommand{\T}[0]{{\rm T}}

\title[Determination and estimation of optimal quarantine duration for infectious diseases]{Determination and estimation of optimal quarantine duration for infectious diseases with application to data analysis of COVID-19}

\author{Ruoyu Wang$^{1,2,}$, and
Qihua Wang$^{1,2,*}$\email{qhwang@amss.ac.cn} \\
$^{1}$Academy of Mathematics and Systems Science, Chinese Academy of Sciences, Beijing 100190, China.\\
$^{2}$University of Chinese Academy of Sciences, Beijing 100049, China.}

\begin{document}


\date{{\it Received September} 2020. {\it Revised December} 2020.  {\it
Accepted February} 2021.}



\pagerange{\pageref{firstpage}--\pageref{lastpage}} 
\volume{0}
\pubyear{2021}
\artmonth{February}


\doi{XXX}


\label{firstpage}


\begin{abstract}
Quarantine measure is a commonly used non-pharmaceutical intervention during the outbreak of infectious diseases.  A key problem for implementing quarantine measure is to determine the duration of the quarantine. Different from the existing methods that determine a constant quarantine duration for everyone, we develop an individualized quarantine rule that suggests different quarantine 
durations for individuals with different characteristics. The proposed quarantine rule is optimal in the sense that it minimizes the average quarantine duration of uninfected people with the constraint that the probability of symptom presentation for infected people attains the given value closing to 1. The optimal solution for the quarantine duration is obtained and estimated by some statistical methods with application to
analyzing COVID-19 data.
\end{abstract}

%



\maketitle


%

\section{Introduction}
During the outbreak of infectious diseases (e.g. EVD, SARS, MERS, and COVID-19), quarantine measures are commonly implemented to limit disease transmission and
morbidity. Extensive research has shown that quarantine is important in reducing the number of people infected and the number of deaths \citep{Lipsitch2003, Ferguson2006}, especially when there is no effective treatment for the disease. See \cite{review2020} for a recent review. To establish a quarantine strategy, some studies use epidemic models such as SEIR (Susceptible-Exposed-Infected-Recovered) type epidemiological models to determine the optimal time-varying quarantine rate by optimal control theory, see for instance \citet{Behncke2000,Yan2008,Ebola2016}. \citet{Lipsitch2003} discussed the relationship between the quarantine fraction of each infectious case’s contacts and the number of person-days in quarantine. However, a key problem when imposing the quarantine measure is to determine the quarantine duration. An extremely long quarantine duration makes sure that most infected individuals would exhibit symptoms under quarantine and then get further quarantine and medical treatment. That is, a long quarantine duration  can stop the virus from spreading to others.
Nevertheless, this may inconvenience uninfected individuals,  incur many extra financial and social costs and even affect  economic development   \citep{quantifying2018}. Hence a good quarantine measure should balance the effectiveness and the cost of the quarantine measure and have a proper duration.

\cite{Farewell2005} proposed to determine the quarantine duration based on the distribution of the incubation period. \cite{Nishiura2009} analyzed the appropriate quarantine period using the quantiles of the incubation period distribution. The existing methods do not consider the characteristics of quarantined individuals and suggest the same quarantine duration for every individual.
Nevertheless, different people may have different probabilities of being infected and different incubation periods of a disease. Indeed, the probability of being infected for every individual is unknown.
However, some individual characteristics  such as 
age, sex, infection rate in the region from  which the individual comes and whether an individual is a close contact,  which may affect the incubation period distribution or the infected probability,  can be observed.	Thus to guarantee the effectiveness and minimize the cost of the quarantine measure, one may intend to set a proper quarantine duration for each potentially exposed individual based on his or her characteristics. To the best of our knowledge, no literature addresses this issue. 

In this paper, we consider the problem and develop  an optimal quarantine rule.
The proposed quarantine rule implements different quarantine durations for different individuals depending on his or her characteristics. We make the rule optimal  by minimizing the average quarantine duration of uninfected people with the constraint that the probability of symptom presentation for infected people attains any given value, which may be close to 1.  We obtain the optimal 
solution for the problem and estimate the optimal solution by some statistical methods.
Coronavirus disease COVID-19 pandemic is known to become a global health crisis since its emergence in Asia late 2019. Considerable attention has been paid to  studying the optimal prevent and control strategy of COVID-19 and  various public health measures such as testing, social distancing, lockdown and quarantine in a macro perspective \citep{Piguillem2020, Charpentier2020,NBERw27102,NBERw26981}. Quarantine is one of the key aspects of infection control during the pandemic of COVID-19. 
This paper focuses on the study of optimal quarantine duration
for infectious diseases with application to data analysis of COVID-19, which is not discussed in all the aforementioned literature. 
Comparing to the standard quantile methods due to \cite{Farewell2005},  \cite{Nishiura2009}
and \cite{liu2020}, 
the data analysis results demonstrate that our method suggests a shorter average quarantine 
duration while keeping the risk of virus spreading  below a given level.  That is, the proposed method
can  keep the risk of virus spreading at the same low level as the standard methods in addition to 
saving cost of days lost. After quarantine, uninfected individuals may work and study by keeping 
some social distance or some other simple and easy measures. Hence, this paper makes a significant   contribution to decreasing   financial and social costs and impact on economic development
with the assurance of controlling the epidemic.  

\section{Optimal quarantine rule}
Let $X$ be a feature vector describing the characteristics of a potentially exposed  individual.
Let $\mathcal{X}$ be the support of $X$ and let $I$ be a variable indicating whether or not the individual has been infected  ($I=1$ if infected and $I=0$ otherwise). Clearly, $I$ is unknown when we decide to quarantine the individual.
A quarantine rule $t(\cdot)$ is a map that maps the feature $X$ to a positive number. Under the quarantine rule $t(\cdot)$, the quarantine duration for an individual with feature value $X=x$ is  determined to be $t(x)$ before quarantine, whether the individual is infected or not. 
An infected individual has a low risk of infecting others if the individual has symptom presentation during the quarantine and hence is not released from the quarantine. A good quarantine duration should ensure a large enough probability that an infected individual has symptom presentation during the quarantine  and minimize the average quarantine duration of uninfected individuals. Let $Y > 0$ be the incubation period of the infectious disease for $I=1$ and the incubation period is not defined for $I=0$. Then the problem of finding the optimal quarantine rule can be expressed as finding a map that minimizes the following problem
\begin{align}\label{orpro}
&\min_{t}\mathbb{E}_0t(X)\quad s.t. \quad 1 - \mathbb{P}_1(Y\leq t(X))\leq \epsilon,
\end{align}
where $\epsilon$ is a predefined small positive number (e.g. 0.05) and the subscript $0$ or $1$ denotes that the expectation or probability is taken conditional on $I=0$ or $1$. For any given quarantine rule $t(\cdot)$, we call $\mathbb{E}_0t(X)$ the average quarantine duration of uninfected people, $\mathbb{P}_{1}(Y\leq t(X))$ the probability of symptom presentation of an infected individual during the quarantine and call $1 - \mathbb{P}_1(Y\leq t(X))$ the escape probability throughout this paper.

If there is no available feature $X$, problem (\ref{orpro}) reduces to
\[\min_{t}t\quad s.t. \quad 1 - \mathbb{P}_1(Y\leq t)\leq \epsilon.\]
This just defines the $1-\epsilon$ quantile of incubation period distribution. In particular, this suggests the 0.95 quantile method due to \cite{Farewell2005} when $\epsilon=0.05$.
\begin{remark}
	Suppose $\theta$ is the proportion of quarantined infected people in all the infected people and $R_0$ is the basic reproductive number of the disease. At the end of quarantine,   the  effective reproductive number  reduces to  $R(\theta,\epsilon,R_0)=(1-\theta)R_0+\theta \epsilon R_0$. If $R(\theta,\epsilon,R_0)<1$, the virus spreading can be controlled. For example, suppose $\theta=0.8$ and $R_0=4$, then the epidemic can be stopped  if we take $\epsilon$ smaller than $1/16$. However, the main purpose of quarantine is to stop the spread of the virus as soon as possible, and hence we usually take $\epsilon$ to be a smaller constant such as $0.05$.
\end{remark}

\subsection{Derivation of the optimal solution}\label{subsec:derive}
Suppose $X=(C,W)$, where $C$ is a categorical variable that takes value in $\{1,\dots,K\}$ and $W\in \mathbb{R}^d$ is a vector of continuous variables. Let $\mu$ be the product of the counting measure on $\{1,\dots,K\}$ and the Lebesgue measure on $\mathbb{R}^d$. Let $f_1(x)$ be the density function of $X$ conditional on $I=1$ with respect to (w.r.t.) $\mu$ and $f_0(x)$ the density function  of $X$ conditional on $I=0$  w.r.t. $\mu$. We use $F_1(y \mid x)$ to denote the distribution function of $Y$ conditional on $X=x$ and $I=1$ and use $f_1(y\mid x)$ to denote the corresponding density function w.r.t. $\mu$. Then problem (\ref{orpro}) can be reformulated as  
\begin{align*}
&\min_t\int t(x)f_0(x)d\mu(x)\quad \\&s.t.\quad  1 - \int F_1(t(x)\mid x)f_1(x)d\mu(x)\leq \epsilon.
\end{align*}

This is a variation problem and not easy to solve in general. However, we find that the solution to this problem is easy to handle under the following conditions. 
\begin{condition}
	$\forall\ x\in \mathcal{X}$, $f_1(y\mid x)>0$ for any $y>0$ and $f_1(y\mid x)$ is continuous with respect to $y$. Moreover,  $f_1(y\mid x)$ is either strictly monotonous
	with respect to $y$ or unimodal and strictly monotonous with respect to $y$ on both of the monotone intervals.
\end{condition}
\begin{condition}
	$0 <\inf_x\mathbb{P}(I=1\mid X=x) \leq \sup_x\mathbb{P}(I=1\mid X=x) < 1$ and
	$\inf_{x}\sup_yf_1(y\mid x)>0$.
\end{condition}
Condition 1 is a mild condition and can be satisfied by many commonly used parameterizations of the incubation period (e.g. Weibull, lognormal, gamma and Erlang distributions). Condition 2 is a mild regular condition. It is not of practical significance to consider the case where $\mathbb{P}(I=1\mid X=x)=0,1$. If we assume for any $x$, the conditional distribution $f_1(y\mid x)=f_1(y\mid \alpha_x, \lambda_x)$ is Weibull distribution with shape parameter $\alpha_x$ and scale parameter $\lambda_x$, then a sufficient condition for $\inf_{x}\sup_yf_1(y\mid x)>0$ is $\sup_{x}\lambda_x < \infty$. 

Before giving the main theorem, we introduce a quantity that is important in the theorem.
By Bayes formula,  
\begin{equation*}
\frac{f_1(x)}{f_0(x)}=\frac{1 - \mathbb{P}(I=1)}{\mathbb{P}(I=1)}\frac{\mathbb{P}(I=1\mid X=x)}{1-\mathbb{P}(I=1\mid X=x)}.
\end{equation*} 
According to Condition 2, we have $\inf_xf_1(x)/f_0(x) > 0$ and
$\inf_x\sup_yf_1(y\mid x)f_1(x)/f_0(x) \geq \inf_x\sup_yf_1(y\mid x)\inf_xf_1(x)/f_0(x) > 0$.
Define \[c^* = \inf_x\sup_y\frac{f_1(y\mid x)f_1(x)}{f_0(x)}.\]
Then we can establish the following theorem.

\begin{theorem}\label{p1}
	For any $0<c\leq c^*$ and $x\in \mathcal{X}$, define $t_c(x)=\sup\{y:f_1(y\mid x)f_1(x)/f_0(x)\geq c\}$.
	Under Conditions 1 and 2,  if $\epsilon$ is small enough such that $\epsilon \leq 1 - \mathbb{E}_1[F_1(t_{c^*}(X)\mid X)]$,  then there is a unique constant $c_0 \in (0, c^*]$ such that $1 - \mathbb{E}_1[F_1(t_{c_0}(X)\mid X)]=\epsilon$ and $t_{c_0}(\cdot)$ is the unique minimum point of problem (\ref{orpro}).
\end{theorem}
The proof of Theorem 1 is given in Web Appendix A.
In what follows, let us make some intuitive explanations for Theorem \ref{p1}.
Our optimal quarantine rule is determined based on the density ratio
\[\frac{f_1(y\mid x)f_1(x)}{f_0(x)},\]
which is like the likelihood ratio in hypothesis testing. See, e.g.,  \cite{Lehmann2005}.
Suppose we need to determine the quarantine duration for an individual with feature value $X=x_0$. Then $f_1(y\mid x_{0})f_1(x_0)/f_0(x_0)$
is a curve of $y$. For a given $c$, we call the set
\[\{y:f_1(y\mid x_0)f_1(x_0)/f_0(x_0)\geq c\}\]
the high-density ratio period. See the following picture for an illustration (Fig. 1).
\begin{figure}[H]
	\centering
	\includegraphics[scale=0.5]{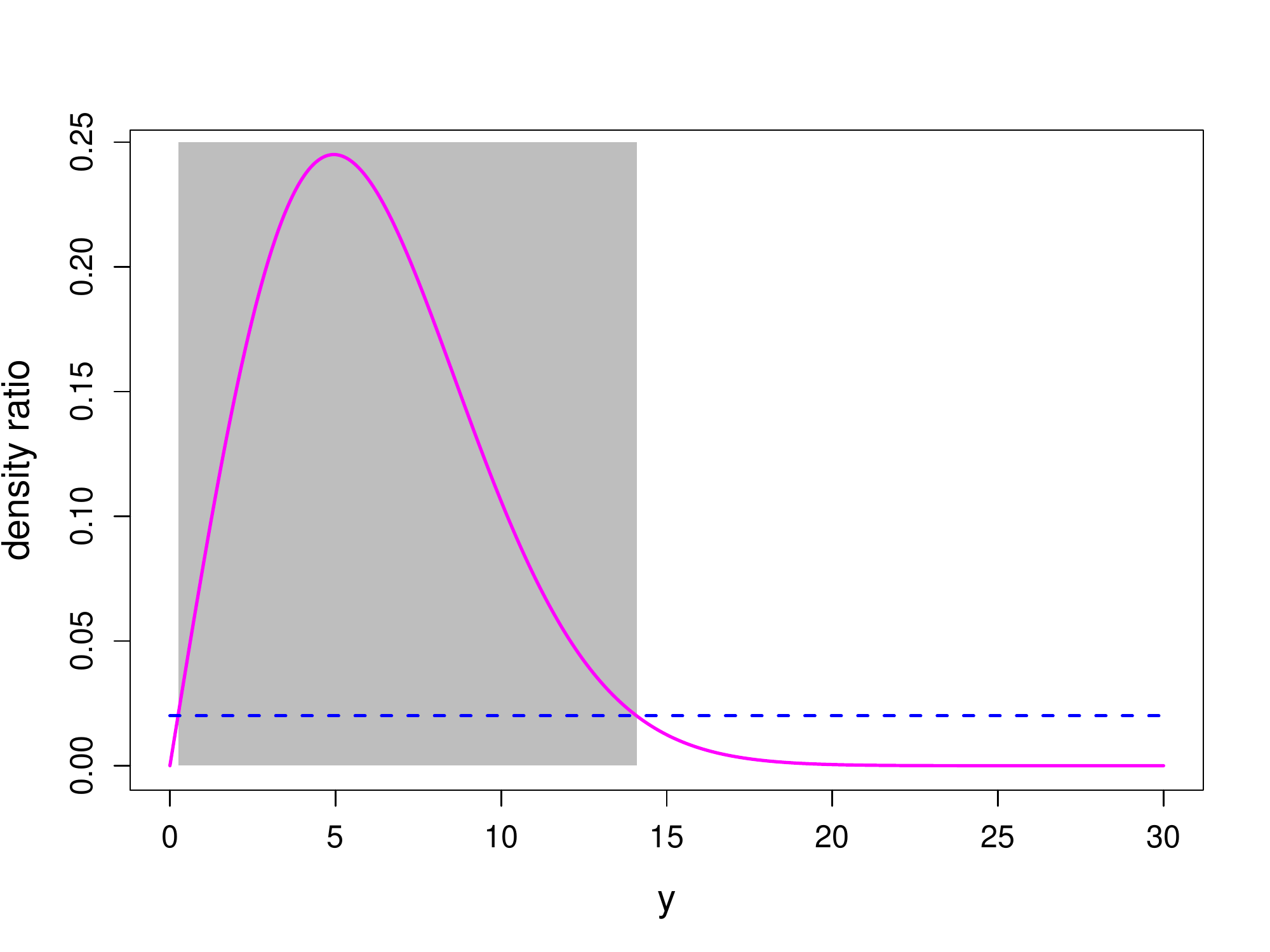}
	\caption{An example for high density ratio period with threshold $0.02$: density ratio: pink solid line; threshold, blue dashed line; high density ratio period, the period between the left and right endpoints of the gray area. This figure appears in color in the electronic version of this article, and any mention of color refers to that version.}
\end{figure}
In the high-density ratio period, the individual has a relatively high probability density of symptom presentation if an individual is infected. A possible quarantine policy is ``release the individual if an individual does not develop any symptom until the end of the high-density ratio period". We denote the resulting quarantine duration by $t_c(x_0)$.
A question is how to determine the threshold value $c$. Clearly, for every $x_0$, $c$ cannot be larger than 
$\sup_y f_1(y|x_0)f_1(x_0)/f_0(x_0)$,  the peak of  the curve. This implies that $c$
cannnot be larger than $c^*$. Condition 1 implies the strict monotonicity and continuity of the escape probability, $1 - \mathbb{E}_1[F_1(t_c(X)\mid X)]$, on $c$. The larger $c$ is, the larger the escape probability is. Thus if $\epsilon \leq 1 - \mathbb{E}_1[F_1(t_{c^*}(X)\mid X)]$, there exists a unique constant $c_0 \in (0, c^*]$ such that $1 - \mathbb{E}_1[F_1(t_{c_0}(X)\mid X)]=\epsilon$. And Theorem 1  states that in this case $t_{c_0}(\cdot)$ is the optimal quarantine rule. 

\begin{remark}
	In practice, the loss of being quarantined for different individuals may also be different. We can easily adapt our framework to this scenario by extending problem (\ref{orpro}) to a more general form
	\begin{align*}
	&\min_{t}\mathbb{E}_0w(X)t(X)\quad s.t. \quad 1-\mathbb{P}_1(Y\leq t(X))\leq \epsilon
	\end{align*}
	where $w(x) > 0$ is a weighting function which indicates different costs of quarantine for different individuals. In this case a modified version of Theorem 1 with $f_0(x)$ in the definition of $t_c(x)$ replaced by $w(x)f_0(x)$ follows directly under Conditions 1 and 2 if $0 <\inf_x w(x) < \sup_x w(x) < \infty$.
\end{remark}

\subsection{Estimation}\label{sec: estimation}

Now we propose an estimation procedure for the optimal quarantine duration for any $x\in\mathcal{X}$.
To estimate the optimal quarantine duration given in Theorem 1, we need to estimate $f_1(y\mid x)$, $f_1(x)$ and $f_0(x)$. Suppose we have historical quarantine data denoted by
$(Y_1,X_1, I_1),\dots,(Y_{n},X_{n}, I_n)$. Note that in contrary to the scenario we considered in Section \ref{subsec:derive}, in the historical data we know whether an individual is infected and this makes our estimation method possible. Here we define $Y_i=0$ 
for samples with $I_i=0$ for $i=1,\dots,n$. Then $f_1(y\mid x)$, $f_1(x)$ and $f_0(x)$ can be estimated 
consistently by either standard parametrical or nonparametrical methods, e.g. maximum likelihood method or kernel smooth method \cite{hansen2008uniform, VanderVaart2000AS}. Suppose $\widehat{f}_1(y\mid x)$, $\widehat{f}_1(x)$ and $\widehat{f}_0(x)$ are the resulting estimators. Then $c^*$ can be estimated by $\widehat{c}^*=\inf_x\sup_y\widehat{f}_1(y\mid x)\widehat{f}_1(x)/\widehat{f}_0(x)$. Let $\widehat{F}_1(y\mid x)=\int_0^y\widehat{f}_1(s\mid x)ds$ be the estimated conditional distribution and $\widehat{t}_{c}(x)=\sup\{y:\widehat{f}_1(y\mid x)\widehat{f}_1(x)/\widehat{f}_0(x)\geq c\}$. Then $c_0$ can be estimated by the solution of 
\begin{equation}\label{eq:ee}
1 - \frac{1}{n_{1}}\sum_{I_i=1}\widehat{F}_1(\widehat{t}_c(X_i)\mid X_i)= \epsilon
\end{equation}
as an equation of $c$ on the interval $(0,\widehat{c}^*]$, where $n_{1} = \sum_{i=1}^{n}I_{i}$ is the number of infected people and $\epsilon$ is a user specified positive number that meets the conditions of Theorem \ref{p1}. The resulting estimator of $c_0$ is denoted by $\widehat{c}_0$. Finally, the estimator of the optimal quarantine duration is $\widehat{t}_{\rm opt}(x)=\sup\{y:\widehat{f}_1(y\mid x)\widehat{f}_1(x)/\widehat{f}_0(x)\geq \widehat{c}_0\}$.
Under some regularity conditions, we show that $\widehat{t}_{\rm opt}(x)$ converges  to the optimal quarantine rule  provided in Theorem \ref{p1} in probability uniformly in $x$. Details on the regularity conditions and the convergence rate of $\widehat{t}_{\rm opt}(x)$ are relegated to Web Appendix B.
\section{Simulation}
\begin{figure}[h]
	\centering
	\subfigure[Scenario 1.]{
		\begin{minipage}[t]{0.45\textwidth}
			\centering
			\includegraphics[scale=0.5]{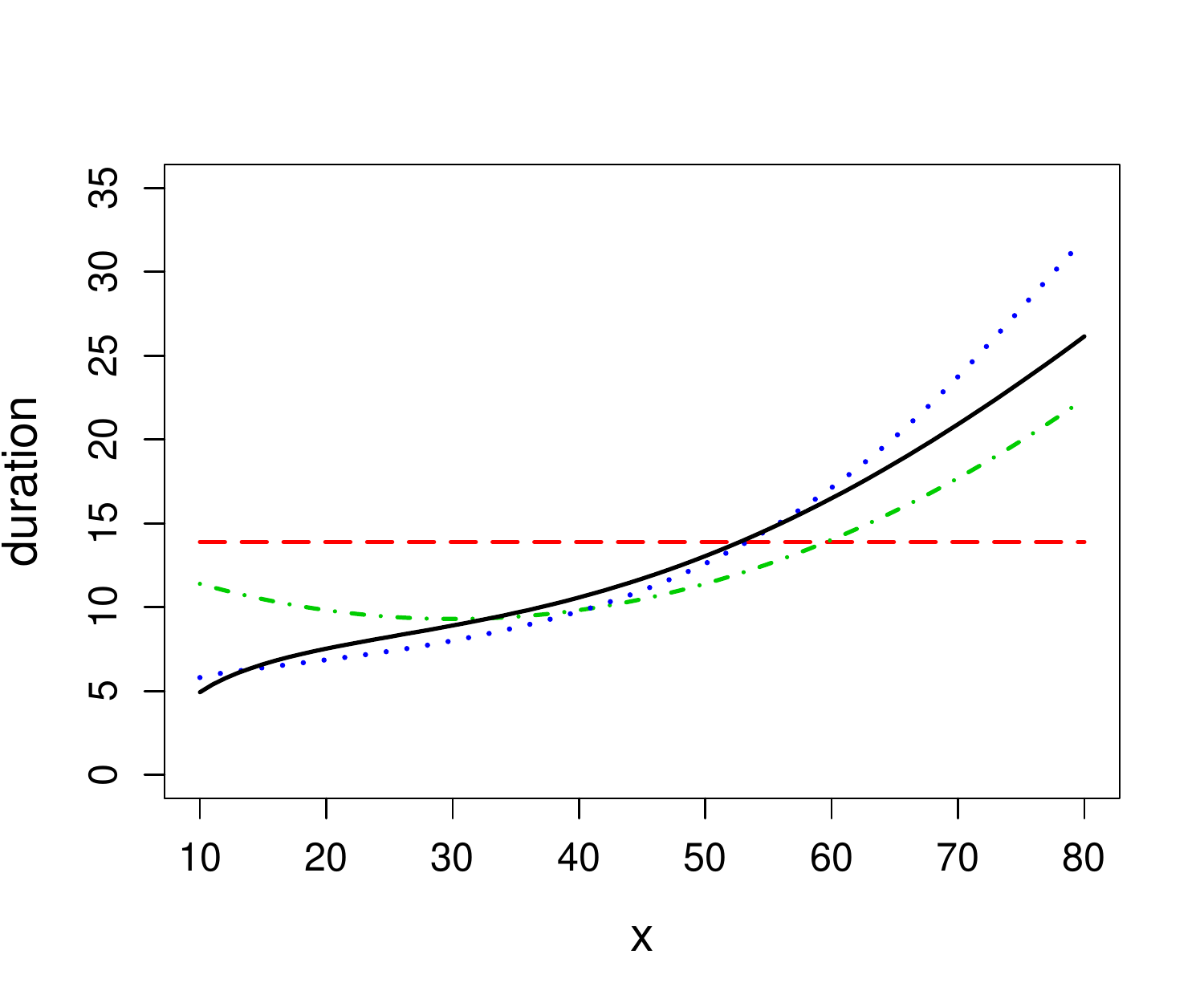}
		\end{minipage}
	}
	\subfigure[Scenario 2.]{  
		\begin{minipage}[t]{0.45\textwidth}
			\centering
			\includegraphics[scale=0.5]{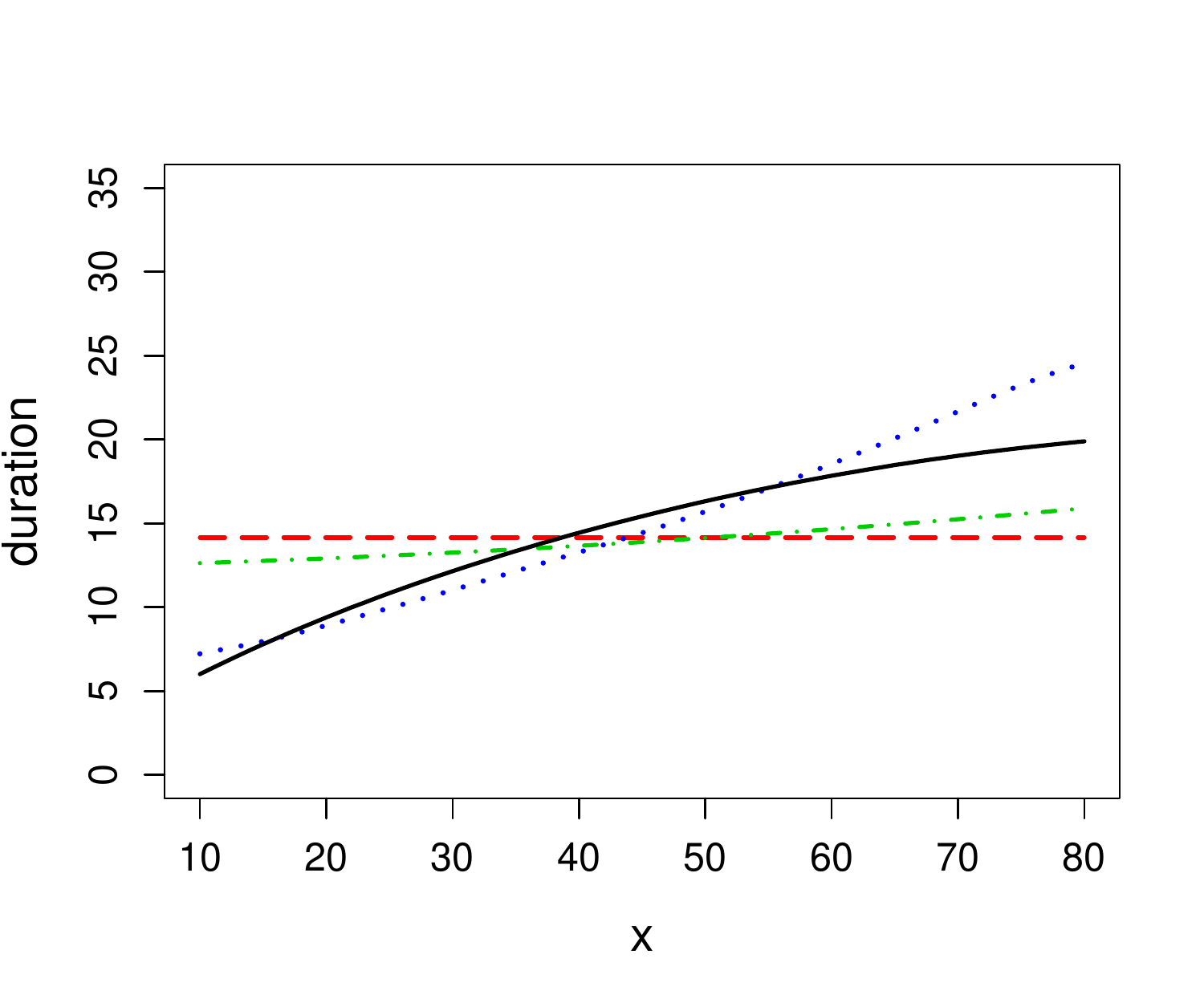}
		\end{minipage}
	}\\
	\subfigure[Scenario 3.]{
		\begin{minipage}[t]{0.45\textwidth}
			\centering
			\includegraphics[scale=0.5]{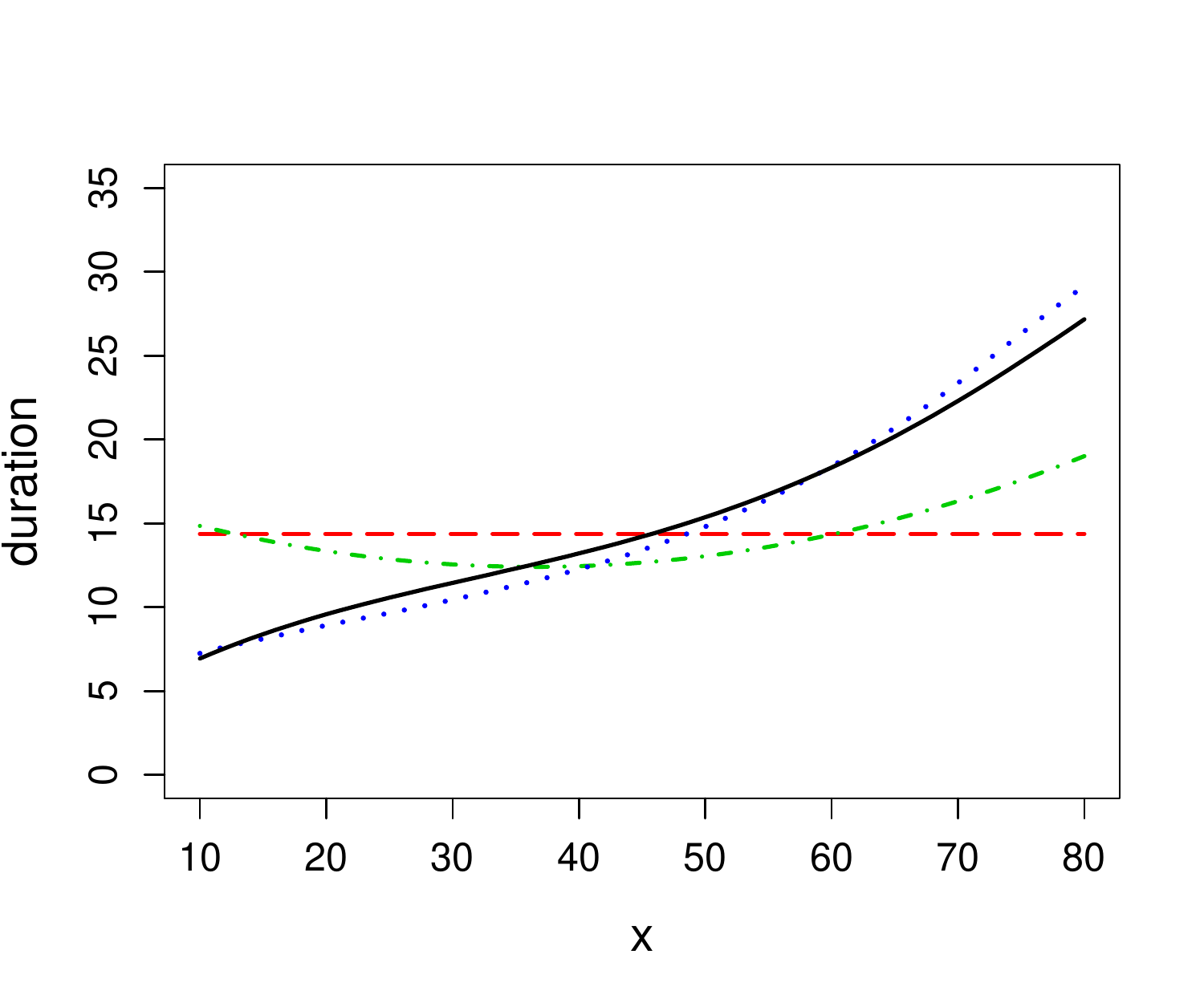}
		\end{minipage}
	} 
	\subfigure[Scenario 4.]{  
		\begin{minipage}[t]{0.45\textwidth}
			\centering
			\includegraphics[scale=0.5]{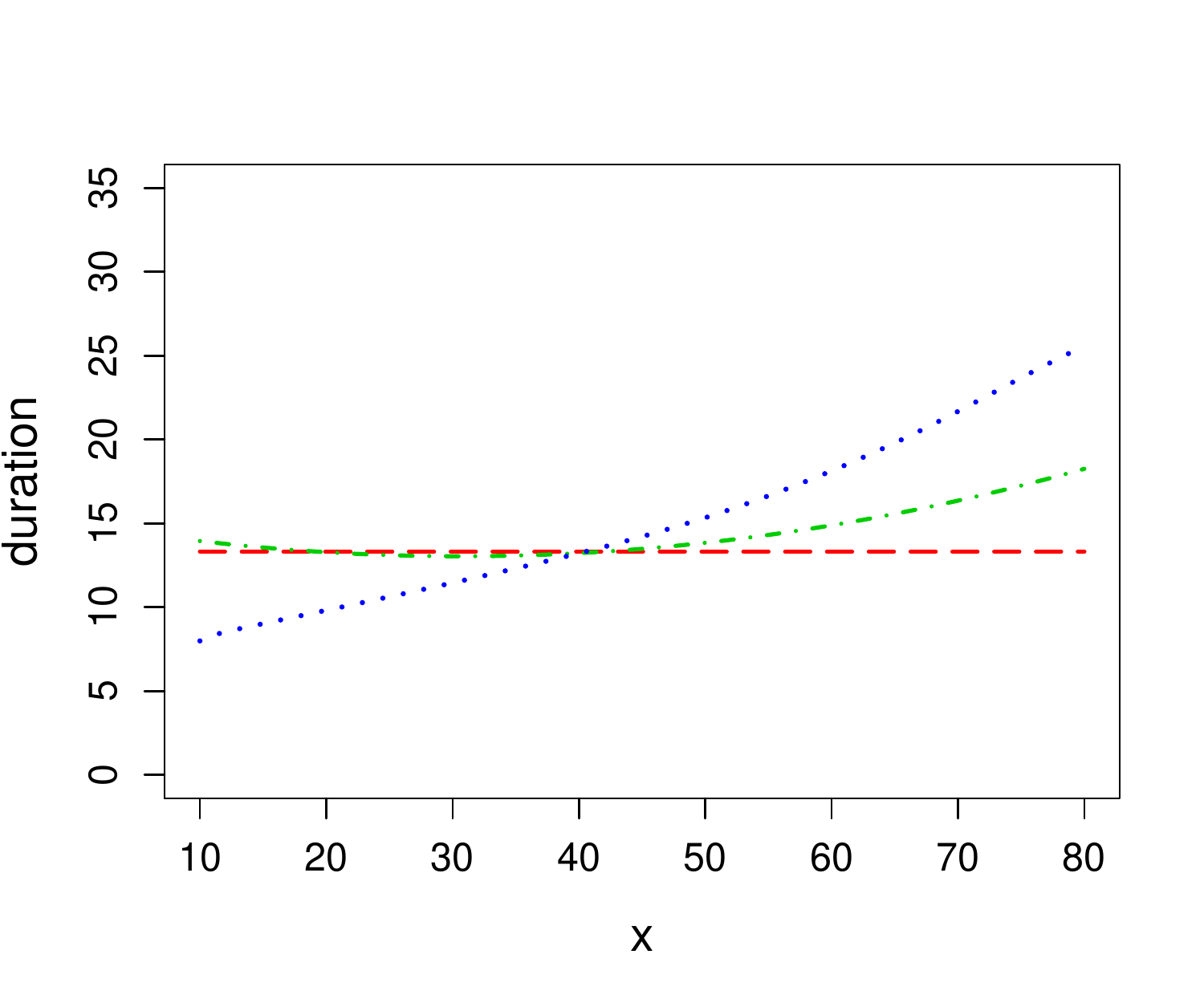}
		\end{minipage}
	}
	\caption{Quarantine duration for people with different feature values: 0.95 quantile, red dashed line; 0.95 conditional quantile, green dashed dotted line; optimal duration, blue dotted line; theoretical optimal duration, black solid line. This figure appears in color in the electronic version of this article, and any mention of color refers to that version.}\label{fig: sim}
\end{figure}

In this section, some simulation studies are conducted to evaluate the performance of the optimal quarantine rule.
Let $\text{TN}(\mu, \sigma^2, a, b)$ be the distribution of a truncated normal variable with 
mean $\mu$ and variance $\sigma^2$  which is truncated to lie in $[a, b]$. First, we generate $(I, X)$ from the following model
\[ I \sim \text{Bernoulli}(0.05), \ X \mid I = 1 \sim \text{TN}(55,625,10,80), \ X \mid I = 0 \sim \text{TN}(25,400,10,80).\]
To evaluate the performance of the optimal quarantine rule under different situations, we consider four data generation processes for the distribution of $Y$ conditional on $X$ and $I = 1$
\begin{itemize}
	\item[] {\bf Scenario 1}: $Y \mid X=x, I = 1 \sim \text{Weibull}(1.5, 4.5 + 0.0025(x - 30)^2)$;
	\item[] {\bf Scenario 2}:  $Y \mid X=x, I = 1 \sim \text{Weibull}(1.5, 3 + \log x)$;
	\item[] {\bf Scenario 3}:  $Y \mid X=x, I = 1 \sim \text{lognormal}(1.5, 0.6 + 0.0002(x - 35)^2)$;
	\item[] {\bf Scenario 4}:  $Y \mid X=x, I = 1 \sim 0.5 * \text{Weibull}(1.5, 4.5 + 0.0025(x - 30)^2) + 0.5 * \text{Weibull}(4, 10)$.
\end{itemize}
In the simulation, we generate $10000$ independent and identical distributed samples from the aforementioned data generation processes. Then
$f_{1}(x)$ and $f_{0}(x)$ are estimated by kernel method. And we assume a Weibull working model for $f_1(y\mid x)$:
\[f_1(y\mid x, \alpha,\gamma)=\frac{\alpha}{\gamma^\T v(x)}\Big(\frac{y}{\gamma^\T v(x)}\Big)^{\alpha - 1}
\exp\Big\{-\Big(\frac{y}{\gamma^\T v(x)}\Big)^{\alpha}\Big\},
\]
where $v(x)=(1, x, x^2)^\T$. The parameters $\alpha$ and $\gamma$ are estimated by the maximum likelihood method. Then we estimate the optimal quarantine rule by the procedure proposed in Section \ref{sec: estimation} with $\epsilon = 0.05$. Under the conditions of Theorem \ref{p1}, \eqref{eq:ee} has a unique solution with probability approaching $1$. However in finite sample the equation may not have a solution. In this case we simply take $\widehat{c}_0 = \widehat{c}^*$ and this treatment performs fairly well in our simulation. We consider the aforementioned four scenarios to evaluate the robustness of the optimal quarantine rule against the violation of model or distribution assumptions.
The working model is correctly specified under Scenario 1; the function form of the scale parameter is misspecified under Scenario 2; the conditional distribution is misspecified under Scenario 3; and the monotonicity assumption is violated under Scenario 4. There are two other ways to make sure $1 - \mathbb{P}_1(Y\leq t(X))\leq 0.05$. One is to omit the feature variables and use the $0.95$ sample quantile of the incubation period as the quarantine duration for everyone \citep{Farewell2005} and another is to use the $0.95$ estimated quantile of the conditional incubation period distribution as the quarantine duration for people with the corresponding feature value \citep{liu2020}. Quarantine durations for people with different feature values obtained by the proposed method and the two quantile methods under the four scenarios are plotted in Fig. \ref{fig: sim}. All the results are averaged over 200 simulation datasets.
We do not plot a theoretical optimal duration in Scenario 4 because the data generation process violates the assumptions of Theorem \ref{p1} under Scenario 4. From Fig. \ref{fig: sim}, we can see that the estimated optimal duration is close to the theoretical optimal duration  when the model is correctly specified, which confirms the convergence result in Web Appendix B. When the model is misspecified, the estimated optimal durations deviate from the theoretical ones. However, the estimated optimal durations still capture some trends of the theoretical optimal durations. Next, we evaluate the performance of the three quarantine rules under different scenarios. We calculate the average quarantine duration (AQD) of uninfected people and the escape probability (EP). The results are summarized in the following table. Because non-integer quarantine duration is  impractical, the quarantine duration is rounded to the nearest integer in calculation. All the results are averaged over 200 simulation datasets.
\begin{table}
	\centering
	\caption{Average quarantine duration of uninfected people and escape probability associated with the three quarantine rules under different scenarios}\label{table: sim}
	\begin{tabular}{cccc}
		\toprule
		Scenario & Method & AQD & EP \\
		\midrule
		\multirow{3}*{1}& \tl{0.95 quantile}& 13.90 & 5.1\%\\
		& \tl{0.95 conditional quantile}& 10.45 & 5.1\%\\
		& \tl{optimal quarantine rule}& 9.33 & 5.1\%\\
		\midrule
		\multirow{3}*{2}& \tl{0.95 quantile}& 14.17 & 5.3\%\\
		& \tl{0.95 conditional quantile}& 13.43 & 5.2\%\\
		& \tl{optimal quarantine rule}& 11.80 & 5.0\%\\
		\midrule
		\multirow{3}*{3}& \tl{0.95 quantile}& 14.39 & 5.0\%\\
		& \tl{0.95 conditional quantile}& 13.20 & 5.4\%\\
		& \tl{optimal quarantine rule}& 11.41 & 5.2\%\\
		\midrule
		\multirow{3}*{4}& \tl{0.95 quantile}& 13.29 & 5.3\%\\
		& \tl{0.95 conditional quantile}& 13.50 & 3.0\%\\
		& \tl{optimal quarantine rule}& 12.17 & 4.7\%\\
		\bottomrule
	\end{tabular}
\end{table}

From Table \ref{table: sim} we can see that the estimated optimal quarantine rule performs well in the aspect of average quarantine duration and escape probability in the four scenarios.  The  estimated optimal quarantine rule does have some robustness against model misspecification and the violation of the monotonicity assumption
although the performance is not as good as that of the case where the model is correctly specified. This may be due to the fact that the optimal quarantine rule combines information contained in $f_1(y\mid x)$, $f_1(x)$ and $f_0(x)$, and the estimated optimal quarantine rule is able to extract information from the marginal feature distributions $f_1(x)$ and $f_0(x)$ even though the conditional distribution model is misspecified. Some extra simulation results with the choice $\epsilon = 0.01$ are relegated to Web Appendix C.

\section{Application to COVID-19 Data}
\subsection{Optimal quarantine rule using age as a feature}
In this subsection, we apply our method to analyzing COVID-19 data. Demographic features such as age, sex, and comorbidities are important in analyzing epidemiological data \citep{Dowd2020}. The incubation period data along with age information are available from the websites of the centers of disease control, or the daily public reports on COVID-19 in 29 provinces in China and are reported by \cite{liu2020}. In this subsection, we use this dataset to construct the optimal quarantine rule using age as the feature $X$. Here we only use the information of patients who are infected before Jan 23th to avoid the biased sampling problem discussed in \cite{liu2020}. The total number of samples is 1770. We use these data to estimate $f_1(x)$ and $f_1(y \mid x)$. 
In the dataset, the proportions of patients younger than 11 and patients older than 80 are very small (1.9\% and 0.6\% respectively). Considering the accuracy of the estimation, we focus on the people aged between 11 and 80 and take these people as the whole population in our analysis (i.e. $\mathcal{X}=\{11,\dots,\ 80\}$).  We apply the kernel method with a Gaussian associate kernel introduced in \cite{Kokonendji2011} to estimate $f_1(x)$.

The reported integer incubation period is regarded as the least integer greater than or equal to the true incubation period. Let $Z=\lceil Y \rceil$ where $\lceil \cdot\rceil$ is the ceiling function. Then the data are regarded as i.i.d. sample from $Z,X \mid I=1$ and denoted by $(Z_1,X_1),\ \dots,\ (Z_n,X_n)$. We assume  conditional on $X=x$ the incubation period $Y$ follows a Weibull distribution, which is commonly used in analyzing incubation period \citep{lauer2020}. And we further assume the conditional density has the form
\[f_1(y\mid x, \alpha_0,\gamma_0)=\frac{\alpha_0}{\gamma_0^\T v(x)}\Big(\frac{y}{\gamma_0^\T v(x)}\Big)^{\alpha_0-1}
\exp\Big[-\Big(\frac{y}{\gamma_0^\T v(x)}\Big)^{\alpha_0}\Big],
\]
where $v(x)=(1, x, x^2)^\T$ and $\alpha_0$ and $\gamma_0=(\gamma_{10},\gamma_{20},\gamma_{30})^\T$ are unknown parameters satisfying
$\alpha_0$ and $\gamma_0^\T v(x) > 0$. Let $\alpha>0$, $\gamma=(\gamma_1,\gamma_2,\gamma_3)^\T$ and $V_i=(1,X_i,X_i^2)^\T$ for $i=1,\dots,\ n$. Then the log likelihood function \cite{ICLik1992} is
\[l(\alpha,\gamma)=\frac{1}{n}\sum_{i=1}^n\log\Big\{\exp\Big[-\Big(\frac{Z_i-1}{\gamma^\T V_i}\Big)^\alpha\Big]-
\exp\Big[-\Big(\frac{Z_i}{\gamma^\T V_i}\Big)^\alpha\Big]\Big\},\]
and $f_1(y\mid x)$ can be estimated by $f_1(y\mid x, \widehat{\alpha},\widehat{\gamma}),$ where $(\widehat{\alpha},\widehat{\gamma}^\T)^\T$ is the maximum likelihood estimator. Here we use a quadratic function to fit the conditional distribution based on the exploratory data analysis.  The estimated values of the parameters with standard error in the bracket are listed as follow:
\begin{table}[h]
	\centering
	\caption{Estimated parameters}
	\begin{tabular}{ccccc}
		\toprule
		Parameter&$\alpha_0$& $\gamma_{10}$& $\gamma_{20}$ & $\gamma_{30}$\\
		\midrule
		Estimation&1.57 (0.03) & 9.09 (0.92) & -0.11 (0.04) & 0.0015 (0.0005) \\
		\bottomrule
	\end{tabular}
\end{table}
In the Web Appendix D, we show that the assumed model fits our data well.

Since the number of infected people in China is relatively small compared to the entire population, we use the age distribution of the entire population of China to estimate the age distribution conditional on $I=0$ and apply the kernel method with a Gaussian associate kernel to estimate $f_0(x)$. 

\begin{figure}[H]
	\centering
	\includegraphics[scale=0.6]{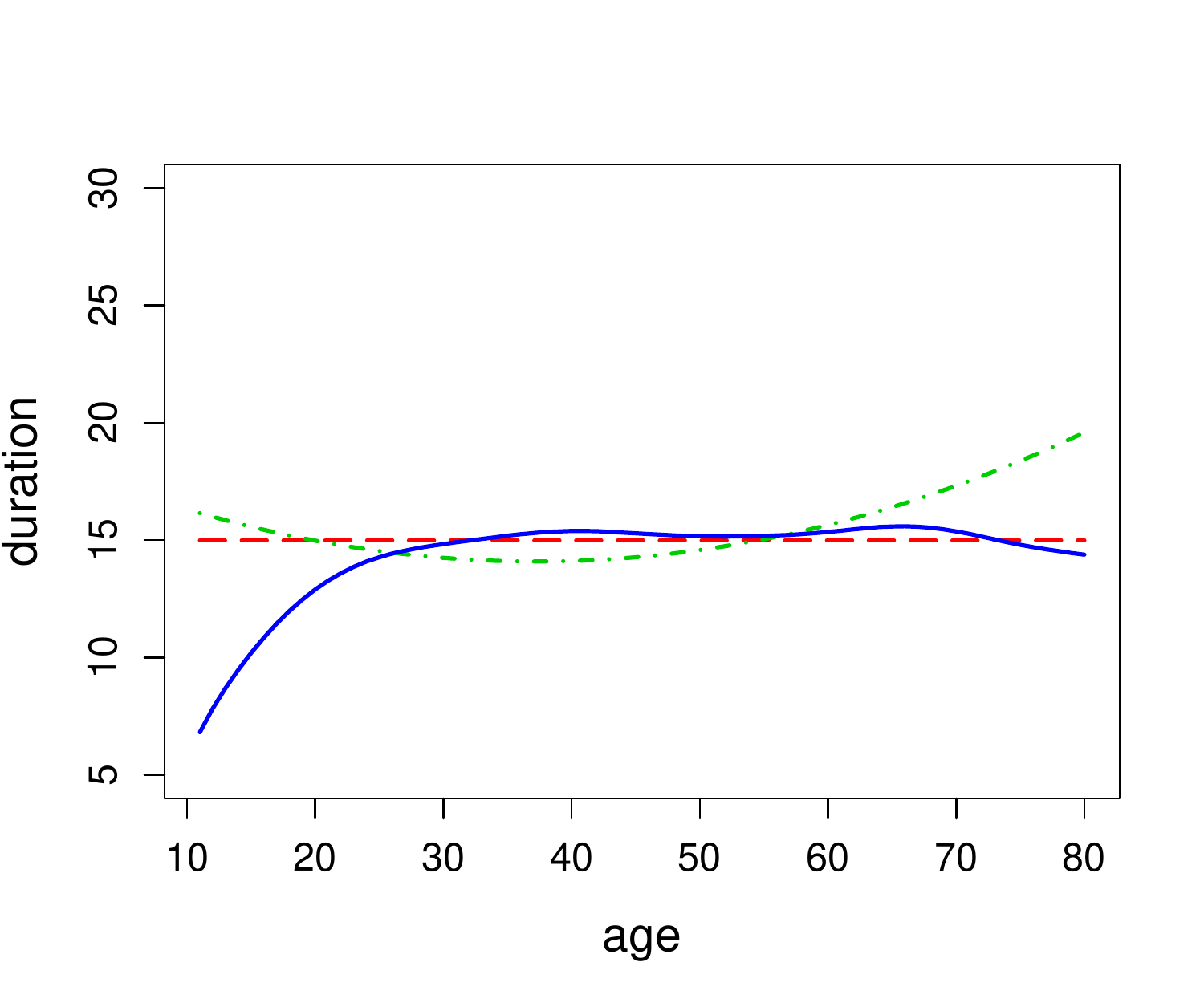}
	\caption{Quarantine duration for people at different ages: 0.95 quantile, red dashed line; 0.95 conditional quantile, green dashed dotted line; optimal duration, blue solid line. This figure appears in color in the electronic version of this article, and any mention of color refers to that version.}\label{qage}
\end{figure}

In this section, we choose $\epsilon=0.05$ which is sufficient to control the epidemic under the scenario discussed in Remark 1.	Quarantine durations for people at different ages obtained by the proposed method and the two quantile methods are plotted in Fig. \ref{qage}.

Figure \ref{qage} shows that the 0.95 sample quantile of incubation period is 15 days, which is one day longer than the current quarantine duration in China. The estimated 0.95 conditional quantile of incubation period of middle-aged people is shorter compared to the young people and the old people. The estimated optimal quarantine durations are close to
15 days for people older than 30 and are shorter than 15 days for people younger than 30. This is because the optimal quarantine rule depends on the probability that an individual is infected and young people is less likely infected in the dataset we consider. For $x\in\mathcal{X}$, 
let $\widehat{t}_{\rm q}(x)$ and $\widehat{t}_{\rm cq}(x)$ be the quarantine durations obtained by $0.95$ sample quantile and $0.95$ estimated conditional quantile, respectively. To compare the performance of these two methods and the optimal quarantine rule, we calculate the average quarantine duration (AQD) of uninfected people and escape probability (EP) by
\[\sum_{j=11}^{80}p_j\widehat{t}_s(j),\]
and
\[1 - \frac{\sum_{i=1}^{1770}1\{Z_i\leq \widehat{t}_s(X_i)\}}{1770},\]
where $p_{j}$ denotes the population proportion aged $j$ in China for $j = 11, \dots, 80$ and $s$ denotes $\rm q, cq$  or $\rm opt$, respectively.
Because non-integer quarantine duration is not practical, the quarantine duration is rounded to the nearest integer in calculation.
The results are listed in Table \ref{table: realdata1}.
\begin{table}[H]
	\centering
	\caption{Average quarantine duration of uninfected people and escape probability associated with the three quarantine rules using age as a feature}\label{table: realdata1}
	\begin{tabular}{ccc}
		\toprule
		Method& AQD & EP\\
		\midrule
		\tl{0.95 quantile} & 15.00 & 3.3\%\\
		\tl{0.95 conditional quantile} & 15.04 & 3.3\%\\
		\tl{optimal quarantine rule} & 14.32 & 3.8\%\\ 
		\bottomrule
	\end{tabular}
\end{table}
Table \ref{table: realdata1} shows that the optimal quarantine rule has the shortest average quarantine duration with guaranteed escape probability. The 0.95 conditional quantile and the optimal quarantine rule are derived based on the conditional distribution model of incubation period. Their reasonable escape probabilities in Table \ref{table: realdata1} also justify our model assumption. The improvement is not great in terms of average quarantine duration. The reason may be that age does not provide sufficient  information for obtaining 
a quarantine rule with good performance. Next, let us consider an example with infection rate in the individual's origin country observed in addition to age. 

\subsection{Optimal quarantine rule based on age and infection rate of origin country}
Travel quarantine for out-of-country travellers and residents from another country is a common policy around the world during the COVID-19 pandemic. When determining quarantine duration, the traveller's age and infection rate of the disease in the origin country can be observed. In this case, infection rate in a traveller's origin country is an important feature that reflects the probability that the traveller is infected. For every country we can calculate a current infection index (CII):  
$\text{CII} = 10^6*a/b$ where $a$ is the number of new cases in the  country during the last two weeks and $b$ is the total population of the country. Here we multiply the rate by a constant $10^6$ to avoid this index being too small. 
We only consider the number of infections in the last two weeks because the number of infections before two weeks provides little information about the infection probability of current traveller. 
We divide the countries with different CIIs into three groups because many countries have similar infection rates. 
Countries with CII larger than 300 are divided into the high risk group, countries with $50<\text{CII}\leq 300$ are divided into the medium risk group and countries with $\text{CII}\leq 50$ are divided into the low risk group. Besides age, we take the risk level of the traveller's origin country as a feature.

In this subsection, we obtain the optimal quarantine rule using information from multiple datasets. We consider 79 countries in our model since their data are relatively complete in all the data sources.
The number of confirmed cases of each country is reported by the Center for Systems Science and Engineering (CSSE) at Johns Hopkins University (JHU) \citep{dong2020interactive}.  We use the number of cases confirmed between May 1st to May 14th in each country to calculate the current infection index. Web Table S2 shows countries at different risk levels.

As in the previous subsection, we focus on the people aged between 11 and 80. We approximate the feature distribution of uninfected people by the distribution of the entire population
(people in the 79 countries) and estimate $f_0(x)$ by the kernel method with a Gaussian associate kernel proposed by \cite{Kokonendji2011} using data from the website of the United Nations \citep{UNdata}. 
Data of 5008 COVID-19 patients from \cite{naturedata} are used to estimate $f_1(x)$. However, we take the proportion of confirmed cases from different countries reported by CSSE at JHU instead of that in the dataset \citep{kraemer2020epidemiological} of \cite{naturedata} since the proportion reported by CSSE at JHU is regarded more exact.

\begin{figure}[H]
	\centering
	\includegraphics[scale=0.6]{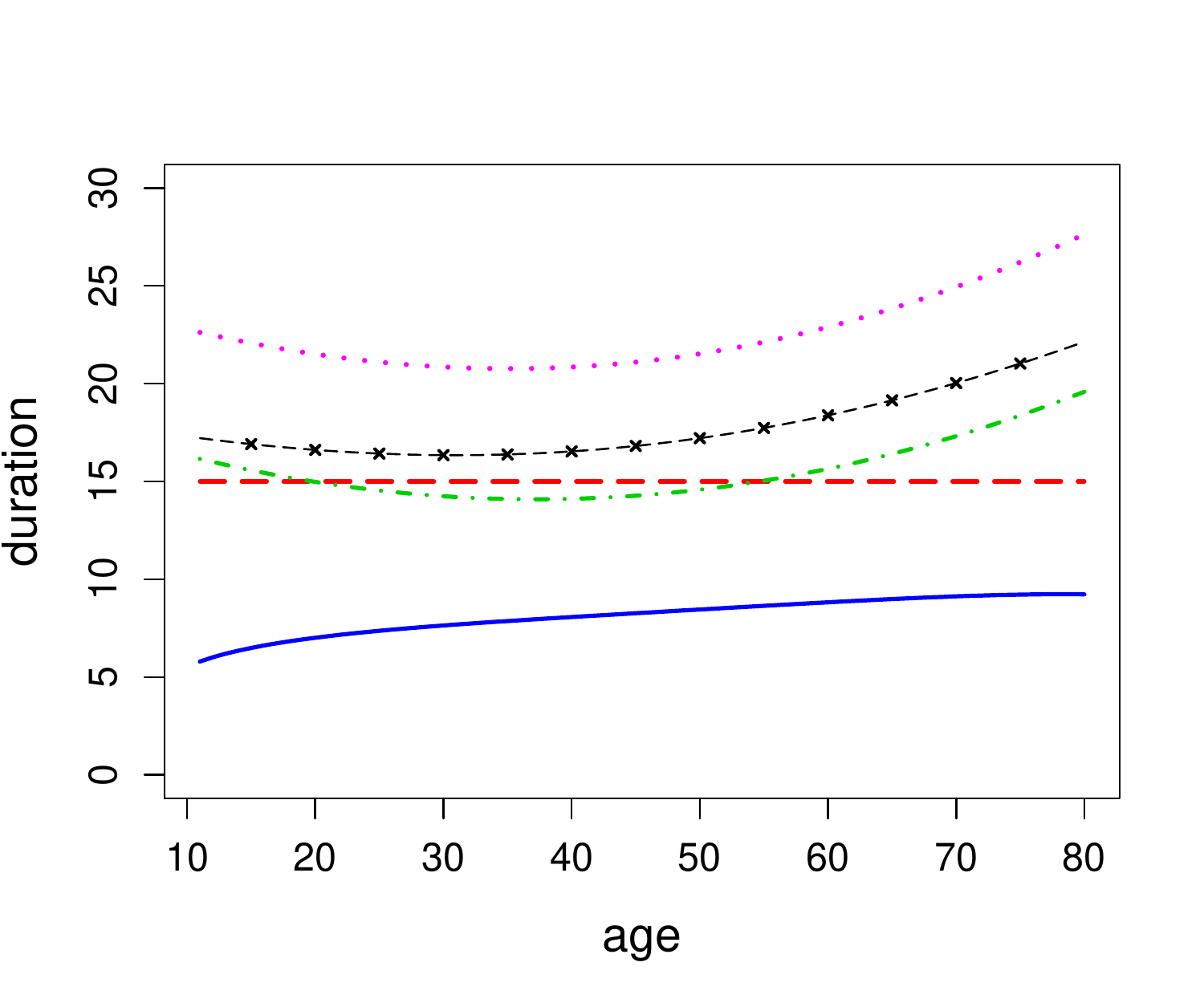}
	\caption{Quarantine durations for people at different ages: 0.95 quantile, red dashed line ; conditional quantile, green dashed dotted line; optimal quarantine duration for travellers from high risk countries, pink dotted line; optimal quarantine duration for travellers from medium risk countries, black short dashed line with crosses; optimal quarantine duration for travellers from low risk countries, blue solid line. This figure appears in color in the electronic version of this article, and any mention of color refers to that version.}\label{qage-risk}
\end{figure}

The dataset of \cite{naturedata} does not contain the incubation period of the patients. To overcome this difficulty, we assume that the distributions of the incubation period for patients at the same age are the same across countries at different risk levels. Thus we can use the conditional distribution model of the incubation period fitted in the previous subsection to impute the missing incubation period. Then we can estimate the three individualized quarantine durations using the imputed dataset. Here we employ the multiple imputation method which is standard in missing data literature. See, e.g., \cite{Rubin2019}. We impute the dataset ten times and average the resulting estimators over different imputed datasets. Quarantine durations  obtained by the sample 0.95 quantile, the estimated 0.95 conditional quantile and the estimated optimal quarantine duration are plotted in Fig. \ref{qage-risk}. 
It can be seen that the optimal quarantine rule gives a much longer duration to travellers from the high risk countries, a duration slightly longer than the 0.95 quantile to travellers from the medium risk countries and a very short duration to  travellers from the low risk countries. Optimal quarantine durations for travellers from high, medium, and low risk countries show different trends on age. The trend of high and medium risk countries is consistent with the trend of the conditional quantile curve. This may be because if the infection rate is relatively high, optimal quarantine duration mainly depends on the incubation period. For travellers from low risk countries, the optimal quarantine rule gives shorter quarantine duration for young people compared to old people. The reason may be that in the low risk countries, infection rate of young people is relatively low. The sample 0.95 quantile  and the estimated 0.95 conditional quantile methods give  quarantine durations that are not dependent on the risk level of the origin country since these two methods are independent of the national infected rate by
definition.

We calculate the average quarantine duration (AQD) of uninfected people and the escape probability (EP) for the three methods by a similar procedure as in the previous subsection. The results are reported in Table \ref{table: realdata2}

\begin{table}[H]
	\centering
	\caption{Average quarantine duration of uninfected people and escape probability associated with the three quarantine rules using age and the risk level of the traveller's origin country as features}\label{table: realdata2}
	\begin{tabular}{ccc}
		\toprule
		Method& AQD & EP\\
		\midrule
		\tl{0.95 quantile} & 15.00 & 5.1\%\\
		\tl{0.95 conditional quantile} & 14.99 & 4.9\%\\
		\tl{optimal duration} & 10.94 & 4.2\%\\ 
		\bottomrule
	\end{tabular}
\end{table}

Table \ref{table: realdata2} shows that our optimal quarantine rule shortens the average quarantine duration of uninfected people greatly with the guaranteed probability of finding the infected individual. Comparing the results in Tables \ref{table: realdata1} and \ref{table: realdata2}, we can see that it is significant to add the risk level of the traveller's origin country as a feature for the optimal 
rule. If one can collect  other features which are associated with the incubation period or the probability that an individual is infected, the optimal quarantine rule may perform even better. 

\section{Discussion}	
Although we mainly discuss COVID-19 in this paper, our method is general and can be applied to establishing optimal quarantine rule for any infectious disease as long as some historical quarantine data are available.
Clearly, the conception ``optimal" depends on the available features. As mentioned before, if there is no available feature, then our optimal quarantine duration reduces to the $1-\epsilon$ quantile of the incubation period. There may be some other features that are useful to determine the quarantine duration. For example,  underlying diseases of an individual and whether an individual is a close contact may also serve as important features. Moreover, it is common that a pathogen test is undertaken before starting quarantine. The test result can also serve as an important feature even though the sensitivity and specificity of the test are not that high. If more features are included, more information is used. However, if we use too many features to construct the optimal quarantine rule, it may be hard to estimate the densities and hence the optimal quarantine duration well. Hence there is a tradeoff. 
It is of great importance to select features which are the most important to determine the quarantine duration and use a few features to construct a quarantine rule that uses information sufficiently. This may be an interesting topic for future works.

The proposed quarantine rule is based on some features. Some of them  are stable across time and  the others may change from time to time. For
example, a country with a high infection rate may have a low infection rate after a few
months. Hence we should use the current feature distribution to build the current quarantine rule.

The  expected number of onward infections may be another useful metric. 
The use of the metic  may lead to another rule.  As pointed out 
by a referee, however, it may be impractical to consider such a metric since 
it is hard to obtain related data and model them.
Another quantity one may want to consider is the subsequent infection, that is, the number of infections caused by infected people who are released from the quarantine. 
This is actually considered by the reproductive number, which is discussed in Remark 1.
According to Remark 1, the quarantine rule proposed in this paper controls the subsequent infection in the average sense by the reproductive number. How to control the subsequent infection more precisely may be an interesting direction for future research.


\section*{Acknowledgements}
This research was supported by the National Natural Science Foundation of China (General
project 11871460 and project for Innovative Research Group in China 61621003), and a
grant from the Key Lab of Random Complex Structure and Data Science, CAS.
\section*{Data Availability Statement}
Age-specific population data of each country are available from the website of United Nations: https://population.un.org/wpp/Download/Standard/CSV. Age information of 5008 COVID-19 patients from different countries is available from the website  https://github.com/ beoutbreakprepared/nCoV2019.
The number of confirmed cases of each country reported by the Center for Systems Science and Engineering (CSSE) at Johns Hopkins University (JHU) can be found on the website  https://github.com/CSSEGISand-Data/COVID-19. All the analyses are
performed with the use of R software, version
3.6.3 (available at http://cran.rstudio.com/bin/windows/base/old/). All the code and data involved in this paper are deposited in Open Science Framework, doi: 10.17605/OSF.IO/5437G.
\bibliographystyle{biom}
\bibliography{inp}
\section*{Supporting Information}
The Web Appendix and Table referenced in Sections 2.1, 3, 4.1 and 4.2 are
available with this article at the Biometrics website on the
Wiley Online Library. All the analyses are
performed with the use of R software, version
3.6.3 (available at http://cran.rstudio.com/ bin/windows/base/old/). All the code and data involved in the supporting information are deposited in Open Science Framework, doi: 10.17605/OSF.IO/5437G.
\label{lastpage}

\end{document}